\documentclass[11pt]{article}
\usepackage{epsfig,amsfonts}
\usepackage[fleqn]{amsmath}
\usepackage{amsthm,amssymb}
\usepackage{graphicx}
\usepackage{hhline}
\usepackage{cite}
\def\be{\begin{equation}}
\def\ee{\end{equation}}
\def\bdm{\begin{displaymath}}
\def\edm{\end{displaymath}}
\def\bea{\begin{eqnarray}}
\def\eea{\end{eqnarray}}

\newcommand{\rd}{\mbox{d}}
\newcommand{\ri}{\mbox{i}}
\newcommand{\re}{\mbox{e}}

\begin{document}

\begin{titlepage}

\begin{flushright}
LPTENS-05/05\\
\end{flushright}

\vspace{.8cm}

\begin{center}
\begin{LARGE}
{\bf Paperclip at $\theta=\pi$}

\end{LARGE}

\vspace{1.5cm}

\begin{large}

{\bf Sergei L. Lukyanov}$^{1,2}$, {\bf Alexei M. Tsvelik}$^3$

\vspace{.2cm}

{\bf and}
\vspace{.2cm}

{\bf Alexander B. Zamolodchikov}$^{1,2,4}$
\end{large}

\vspace{.8cm}

{${}^{1}$ NHETC, Department of Physics and Astronomy\\
     Rutgers University\\
     Piscataway, NJ 08855-0849, USA\\

\vspace{.2cm}

${}^{2}$ L.D. Landau Institute for Theoretical Physics\\
  Chernogolovka, 142432, Russia

\vspace{.2cm}

${}^{3}$ Department of Physics, Brookhaven National Laboratory\\
Upton, NY 11973-5000, USA\\

\vspace{.2cm}
and
\vspace{.2cm}

${}^{4}$ Chaire Internationale de Recherche Blaise Pascal\\
Laboratoire de Physique Th${\acute {\rm e}}$orique
de l'Ecole Normale Sup${\acute {\rm e}}$rieure\\
24 rue Lhomond, Paris Cedex 05, France\\
}
\vspace{.4cm}

\end{center}

\begin{center}
\centerline{\bf Abstract} \vspace{.8cm}
\parbox{11cm}{
We study the ``paperclip'' model of boundary interaction with the
topological angle $\theta$ equal to $\pi$. We propose exact
expression for the disk partition function in terms of solutions
of certain ordinary differential equation. Large distance asymptotic
form of the partition function which follows from this proposal makes
it possible to identify the infrared fixed point of the paperclip
boundary flow at $\theta=\pi$.}
\end{center}

\vspace{0.1cm}

\begin{flushleft}
\rule{4.1 in}{.007 in}\\
{January 2005}
\end{flushleft}
\vfill

\end{titlepage}
\newpage

Field theory models where 
interactions emerge as a consequence 
of geometric constrains imposed 
on the fields are much interest. They are esthetically 
attractive, and they  
often have rich content and important applications.
Typical models of this class are nonlinear sigma models 
where the constraints are imposed in the bulk of the 
space-time. Lately the other type of models, where the 
constraints are imposed only on a boundary, attracts much
attention. Two-dimensional models of this type emerge 
naturally in string theories and in some condensed matter 
problems. In string theories they provide the world-sheet
description of ``branes'', while in condensed matter theory
they  describe either quantum 
impurities or ``quantum dots''.

\begin{figure}[ht]
\centering
\includegraphics[width=10.5cm]{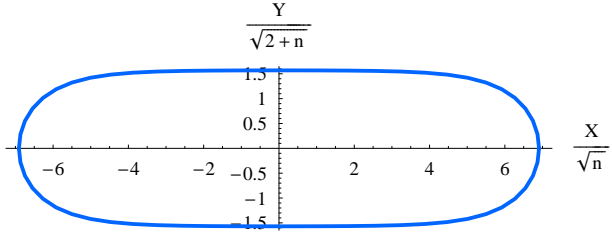}
\caption{The paperclip shape for $r=2\times 10^{-3}$. }
\label{fig-pl}
\end{figure}

In this paper we study one model of this type.  
It is the so-called paperclip model introduced in
\cite{LVZ}. This two-dimensional model of quantum field theory
involves two-component Bose field ${\bf X}(z, {\bar z})= \big(X(z,
{\bar z}), Y(z, {\bar z})\big)$ living  on the disk of radius
$R$. In the bulk, i.e. at $|z| <R$, the field ${\bf X}(z, {\bar z})$ is a free
massless field, as described by the bulk action
\bea\label{baction}
{\cal A}_{\rm bulk}[{\bf X}]
= {1\over\pi}\,\int_{|z|< R}\rd^2z\ \,\partial_{z}{\bf
  X}\cdot \partial_{\bar z}{\bf X}\, ,
\eea
while the boundary values ${\bf X}_B$ of this field, ${\bf X}_B = {\bf
  X}|_{|z|=R}$, are subjected to a nonlinear constraint
\bea\label{bconstaint}
r\,\cosh\big( {\textstyle{X_B\over\sqrt{n}}}\big) -
\cos\big(  {\textstyle{Y_B\over\sqrt{n+2}}}\big)= 0 \,, \ \ \qquad
|Y_{B}| \leq {\pi\over 2}\, \sqrt{n+2}\ .
\eea
Here $n$ and $r$ are real and positive parameters (despite the
notation, $n$ is not necessarily  integer). The renormalization does not affect
the parameter $n$, which is thus a scale-independent constant, while
$r$ ``flows'' under the RG transformations; up to two loops, the flow
is described by the equation
\bea\label{rgflow}
\kappa = (n+1)\  (1 - r^2)\, r^{n}\ ,
\eea
where $\kappa = \textstyle{E_{*}\over E}$ is inversely proportional to the
RG energy scale $E$. Here $E_{*}$ is the integration constant of
the RG equation, which sets up the ``physical scale'' in the model.
As in \cite{LVZ}, we will always choose $E$  equal to $R^{-1}$, the inverse
radius of the disk, so that
\bea\label{kappadef}
\kappa = E_{*} R\,.
\eea
The equation\ \eqref{bconstaint}\ defines a closed curve in the $(X_B, Y_B)$
plane, which at sufficiently small $r$ has a paperclip shape (see
Fig.\,\ref{fig-pl}), hence the name of the model.

When $n$ goes to infinity and simultaneously $r$ goes to $1$ as $r
= 1 - {1\over{2 (n+1) g }}$ with finite $g$, the paperclip curve
becomes a circle 
\bea\label{cbrane} X_{B}^2 + Y_{B}^2 = { {1\over
g}}\ , \eea 
and the flow equation \eqref{rgflow} reduces to
\bea\label{cflow} \kappa = g^{-1}\ \re^{-{1\over{2g}}}\ . 
\eea 
The
boundary constraint \eqref{cbrane} defines the ``circular brane''
model \cite{SLAZ}. 

The ``circular brane'' model has important application in condensed 
matter physics. It is equivalent to the model of dissipative quantum 
mechanics known as Ambegaokar-Eckern-Sch\"on model \cite{AES}. The 
latter is used to describe the low-energy sector of the model of a 
``weakly blockaded'' quantum dot introduced in 
\cite{FurMat95, FFurMat95} (see also the review article \cite{review}).
That model describes an almost open dot which has a large number of 
energy levels and a large charging energy. The dot is connected to 
the bulk via $n$ degenerate channels, and the Ambegaokar-Eckern-Sch\"on 
model applies when the number of channels is very large. In this
application the  circumference $2\pi R$ of the disk in 
\eqref{baction} has obvious interpretation as the inverse temperature
$1/T$, while the large charging energy provides explicit 
ultraviolet cut-off for the model, and the topological angle of 
the circular brane (introduced later in the text) is related to 
the gate voltage. Although the Ambegaokar-Eckern-Sch\"on model 
appears only in the $n\to\infty$ limit of the model 
\eqref{bconstaint} when the paperclip becomes the circle 
\eqref{cbrane}, we have reasons to believe that the paperclip model 
with finite integer $n$ has certain relation to the quantum dot 
model with $n$ open degenerate channels. We intend to explore 
this possible relation in separate work.        

The primary object of interest is the boundary state $|\, B\, \rangle$, in
particular its overlap with the Fock vacuum $|\, {\bf P}\, \rangle$
characterized by the zero-mode momentum ${\bf P} = (P,Q)$ of
the free field ${\bf X}=(X,Y)$. The overlap
\bea\label{zdef}
\langle\, {\bf  P}\, |\,  B\,  \rangle =
Z(\, {\bf P}\, |\, \kappa\, )
\eea
can be expressed through the disk one-point function
\bea\label{fint}
 R^{1/3-{{\bf P}^2/ 2} }\ Z(\, {\bf P}\, |\, \kappa\, )
=\big\langle\,  \re^{\ri {\bf P}\cdot {\bf X}} (0,0)
\, \big\rangle_{\rm disk}=
\int\,{\cal D}{\bf X}\ \re^{\ri\,{\bf
    P}\cdot {\bf X}(0,0)}\ \re^{-{\cal A}_{\rm bulk}[{\bf X}]}
\,,
\eea
where the functional integration is over all fields ${\bf X}(z,
{\bar z})$ obeying the boundary constraint\ \eqref{bconstaint}.

Let us repeat again that Eq.\,\eqref{rgflow}, as well as the
paperclip equation\ \eqref{bconstaint}, was obtained
perturbatively, in the two-loop approximation. Therefore it provides
useful description of the boundary condition only in the weak
coupling regime, where the curvature of the paperclip curve
\eqref{bconstaint} is small everywhere; this requires $n$ to be large,
$n\gg 1$, and $r$ to be sufficiently small (so that $r^n \ll 1$);
according to \eqref{rgflow}, the
last condition is fulfilled at sufficiently small $R$, therefore
Eqs.\,\eqref{bconstaint},\,\eqref{rgflow} provide
ultraviolet (UV) description of the
boundary condition. At large distances (large $R$) and at $n \sim 1$
the higher loops\footnote{The higher loop corrections to
\eqref{bconstaint},\,\eqref{rgflow} are scheme dependent.
There are reasons to believe that 
Eqs.\,\eqref{bconstaint},\,\eqref{rgflow}
are perturbatively exact,
i.e. a scheme exists in which these equations are exact to all orders
in the loop expansion, see \cite{LVZ}.} and non-perturbative corrections
are important. Generally, one expects that at $R\to\infty$ the
paperclip boundary theory ``flows'' to some infrared (IR) fixed point,
and so the $R=\infty$ limit of the boundary state $|\, B\, \rangle$ is
described in terms of conformal boundary theory associated with
the IR fixed point.

In regard to the non-perturbative effects, it is important to realize
that under general definition the paperclip model may involve an additional
parameter, the topological angle $\theta$.
Since topologically the paperclip curve \eqref{bconstaint} is a circle, the
configuration space for the field ${\bf X}(z,{\bar z})$ consists of
topological sectors, each characterized by integer $w$ which is the
number of times the boundary value ${\bf X}_B$ winds around the
paperclip curve when one goes around the disk boundary $|z|=R$.
This allows one to add the weight factors $\re^{{\rm i} w\theta}$
to all contributions to the functional integral coming from the
sectors with the winding number $w$. Thus, in general
\bea\label{topsum}
Z(\, {\bf P}\, |\, \kappa\, )= \sum_{w=-\infty}^{\infty}\re^{{\rm i}w\theta
}\ Z^{(w)}(\, {\bf P}\, |\, \kappa\, )\,,
\eea
where $Z^{(w)}$ receives contributions from the topological sector $w$
only.  Of course, the contributions of the
instanton sectors $w\neq 0$ are invisible in the perturbation theory,
and therefore the UV limit $R\to 0$ of the theory is insensitive to
the topological angle (this statement is not limited to the case
$n\gg 1$; the instanton contributions are suppressed by the powers of
$\kappa$ at any $n$, see \cite{LVZ}). However, the IR  behavior of the
theory can depend on $\theta$ in a significant way.

The paperclip model is in a close analogy with the so called
``sausage'' sigma model proposed and studied in \cite{sausage}
(perhaps better known is the symmetric limit of the sausage model,
that $O(3)$ nonlinear sigma model; its counterpart is the circular brane
model \eqref{cbrane}). The IR physics of the sausage
(and its  $O(3)$ limit) sigma model
strongly depends on its topological angle $\theta$ \cite{Haldane,
Afl}. The
sausage sigma model  is believed to be
integrable at two values of the topological angle, $\theta=0$ and
$\theta=\pi$. The arguments were given in\ \cite{sausage},  where
exact solutions of the sausage model at these two values of
$\theta$ were proposed. According to this proposal, the sausage
model at $\theta=0$ is massive and its solution is described in
terms of factorizable $S$-matrix of three massive particles. On
the contrary, the sausage model at $\theta=\pi$ ``flows'' in the
IR limit to a critical fixed point, which is the $c=1$ CFT of a
compactified free boson; the ``flow'' is described in terms of
certain massless Thermodynamic Bethe Ansatz (TBA) system\
\cite{sausage}. Solutions for the $O(3)$ sigma model at $\theta=0$
and $\theta=\pi$ were previously proposed in \cite{ZZ} and
\cite{ZZA}. By the analogy, one may expect that the paperclip
model is also integrable in two cases $\theta=0$ and $\theta=\pi$.

\begin{figure}[h]
\centering
\includegraphics[width=9cm]{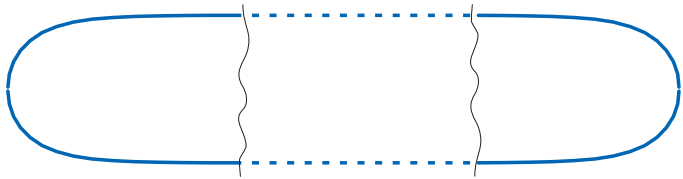}
\caption{The paperclip formed by a junction of two hairpins.}
\label{fig-ampla}
\end{figure}

There is more then just this analogy to suspect that the
paperclip model might be integrable at least at some values of
$\theta$. In \cite{LVZ} a set of commuting local integrals of motion
$\{\, {\mathbb I}_{2l-1}\,\}_{ l=1}^{\infty}$ of the free Bose
theory \eqref{baction}
was displayed which has all
symmetries of the paperclip \eqref{bconstaint}, and moreover it was found
that given the symmetries the set is essentially unique. The idea that
this series has something to do with the paperclip model can be
supported by the analysis of the UV properties of the model.
In the UV limit $R\to 0$
the parameter $r$ in \eqref{bconstaint} becomes small, and the paperclip
grows long in the $X$ direction; in this limit the paperclip can
be regarded as the composition of left and right ``hairpins''
(see Fig.\,\ref{fig-ampla}).

\noindent
By the ``hairpins'' we understand the curves
\bea\label{hairpins}
 {\textstyle{r\over 2}}\  \exp\big(\pm {\textstyle{X_B\over\sqrt{n}}}\big)=
\cos\big(  {\textstyle{Y_B\over\sqrt{n+2}}}\big)\ , \eea where the
sign plus (minus) has to be taken for the left (right) hairpin. If
any one of the curves \eqref{hairpins} is taken as the boundary
constraint, the associated left and right ``hairpin models'' are
conformally invariant, and moreover each has extended conformal
symmetry with respect to certain $ W$-algebra \cite{LVZ}. Although
the $ W$-algebras of the right and left hairpin models are
isomorphic, the generators of $ W^{(\subset)}$ and $
W^{(\supset)}$ are realized by different operators in the space of
states of the free boson theory\ \eqref{baction}. However, the
sets of operators $ W^{(\subset)}$ and $ W^{(\supset)}$ have
nontrivial intersection, which is exactly the set of commuting
integrals $\{{\mathbb I}_{2l-1}\}$\  \cite{LVZ}. For these reasons
we refer to the set $\{{ \mathbb I}_{2l-1}\}$ as the ``paperclip
series'' of local IM.

In \cite{LVZ} an exact expression for the amplitude
$Z_{\theta=0}(\, {\bf P}\, |\, \kappa\, )$ was proposed in terms
of solutions of certain linear differential equation. The proposal
was inspired by remarkable observation of Dorey and Tateo\
\cite{toteo} who found that in somewhat simpler integrable model
of boundary interaction (the minimal CFT with non-conformal
boundary perturbation) the overlap amplitude analogous to
\eqref{zdef} is related to certain monodromy coefficients of the
Schr{$\rm {\ddot o}$}edinger equation with the potential
$|x|^{2\alpha}$. Since \cite{toteo}, this finding was confirmed
and extended to several other integrable models of boundary
interaction\ \cite{blzz,baza,Tsvelik}. Although it is fair to say
that true roots of this relation remain mysterious to us, the
examples suggest that the relation may be rather general. Given a
model of boundary interaction suspect of being integrable, it is
worth trying to identify associated ordinary differential
equation. The proposal of \cite{LVZ} was made according to this
strategy. Let us briefly summarize it here.

The main ingredient is  
the ordinary differential equation
\bea\label{diffa} \bigg[ -{{\rd^2}\over{\rd x^2}}   -
{\textstyle{nP^2\over 4}}\, { {\re^x}\over{1+\re^x}} -
{\textstyle{(n+2)Q^2-1\over 4}} \, {{\re^x}\over{(1+\re^x)^2}} +
\kappa^2\,\big(1+ \re^x\big)^n \bigg]\, \Psi(x) = 0\, , \eea 
where
$P$ and $Q$ are the $X$ and $Y$ components of the zero-mode
momentum ${\bf P}$, and $\kappa$ is the same as in\
\eqref{kappadef}; according to the definition \eqref{kappadef} we
assume for the moment that $\kappa$ is real and positive.

Eq.\,\eqref{diffa} has a form of a stationary zero energy Schr${\rm \ddot
o}$edinger equation with specific potential $V(x)$ given by the
last three terms in \eqref{diffa}. The potential
$V(x)$ is positive and grows fast at large positive $x$, therefore
\eqref{diffa} has a solution $\Xi (x)$ decaying  at
$x\to+\infty$; this condition specifies $\Xi(x)$ uniquely up to
normalization\footnote{Here we slightly change notations for the
solutions of\ \eqref{diffa} as compared to \cite{LVZ}.}. To fix
the normalization, we assume that \bea\label{ksidef} \Xi(x) \to
\kappa^{-{1\over 2}}\
 \exp\bigg\{ -\big({\textstyle{n\over 4}}+\kappa\big)\, x- \kappa\,
\int_0^{\re^x} {\rd u\over u}\, \big((1+u)^{n\over 2}-1\big)\, \bigg\}
\eea as $ x\to+\infty$. On the other hand, $V(x)$ approaches
positive constant $\kappa^2$ at large negative $x$. Hence
Eq.\,\eqref{diffa} has a solution which decays at large negative
$x$; we denote this solution $\Psi_{+}(x)$. The condition,
\bea\label{psiplusdef} \Psi_{+}(x) \to {{\re^{\kappa
x}}\over{\Gamma(1+2\kappa)}} \quad {\rm as} \quad x\to\infty\, ,
\eea
specifies the solution $\Psi_{+}(x)$ uniquely, including its
normalization. Then 
\bea\label{zet0} Z_{\theta=0}(\, {\bf P}\, |\,
\kappa\, ) = {\rm g}_{D}^2\,\sqrt{\pi}\ \Big({{2\kappa}\over
\re}\Big)^{2\kappa}\,W\big[\Xi, \Psi_{+}\big]\,. \eea 
Here and
below ${\rm g}_D = 2^{-1/4}$ is the $g$-factor\ \cite{affleck}\ of
the Dirichlet boundary,
and $W[F,G]$ denotes the Wronskian $F(x)G'(x)-F'(x)G(x)$.
Eq.\,\eqref{zet0} is the proposal of \cite{LVZ}.

It was shown in\ \cite{LVZ} that \eqref{zet0} exhibits the UV (i.e
$\kappa\to 0$) behavior completely consistent with what one
expects from the ``hairpin decomposition'' of the paperclip at
$R\to 0$. At small $\kappa$ the expression\ \eqref{zet0} can be
written as (here we write explicitly the components of ${\bf P} =
(P,Q)$) \bea\label{uvzeto} Z_{\theta=0}(\, P,Q\, |\,\kappa\,) =
B(P,Q)\ F_{\theta=0}(P,Q\, |\,\kappa) + B(-P,Q)\
F_{\theta=0}(-P,Q\, |\,\kappa)\,, \eea where \bea\label{hairamp}
&&B (P,Q) = {\rm
g}_D^2\ \Big({\kappa\over n}\Big)^{\ri  {P\over \sqrt n} }\times\\
\nonumber && \ \ \ \ {{\sqrt{n}\ \Gamma(-\ri\sqrt{n}\,P)\,
\Gamma(1 - {{\ri P}/\sqrt{n}})}\over{\Gamma\big({1\over 2} -
 \sqrt{n+2}\ {Q\over 2}- \ri\sqrt{n}\ {P\over 2})\,
\Gamma\big({1\over 2} +  \sqrt{n+2}\ {Q\over 2}-  \ri\sqrt{n}\
{P\over 2} \big)}}\ , \eea and $F_{\theta=0}(P,Q\, |\, \kappa)$
(apart from the factor $\kappa^{2\kappa}$) admits asymptotic expansion
in a double series in powers of $\kappa$ and $\kappa^{2\over n}$ (here
and below we use the symbol $\simeq$ to indicate relations which hold
in the sense of asymptotic series),
\bea\label{fexpan} F_{\theta=0}(P,Q|\kappa) \simeq
\kappa^{2\kappa}\,\sum_{i,j=0}^{\infty}\,f_{i,j}(P,Q)\,\kappa^{i+{{2j}\over
n}}\,, \eea with $f_{0,0}=1$. From\ \eqref{zet0}, the powers
$\kappa^{2\over n}$ appear from the $\kappa\to 0$ expansion of the
solution $\Xi(x)$, while the integer powers of $\kappa$ come from
the expansion of $\Psi_{+}(x)$.

The expansion\ \eqref{uvzeto} is in good agreement with expected
form of\ \eqref{fint} in the domain $\kappa \ll 1$. Recall that
according to \eqref{kappadef} and \eqref{rgflow} this domain
corresponds to the UV limit of the paperclip model, where the
paperclip \eqref{bconstaint} can be regarded as the composition of
two hairpins, as in Fig.\,\ref{fig-ampla}. Roughly speaking, the two
terms in \eqref{uvzeto} correspond to contributions from the right
and the left hairpins, respectively. More precisely, at small
$\kappa$ and $\Im m\,P\neq 0$ the saddle points of the functional
integral \eqref{fint} correspond to the field configurations where
the boundary values $(X_B, Y_B)$ are mostly concentrated near the
left or the right end of the paperclip \eqref{bconstaint}. The
factors $B(P,Q)$ and $B(-P,Q)$ are precisely the boundary overlap
amplitudes of the right and left hairpin models \eqref{hairpins},
respectively. The powers of $\kappa^{2\over n}$ in \eqref{fexpan}
are associated with the perturbative corrections, due to ``small''
fluctuations around the saddle points, which feel only the small
deviations of the shape of the paperclip \eqref{bconstaint} from
the respective hairpin, while the integer powers of $\kappa$ are
associated with the contributions of the $w\neq 0$ topological
sectors in \eqref{topsum} (see Ref.\,\cite{LVZ} for more detailed
discussion, which also includes the explanation of the factor
$\kappa^{2\kappa}$ in \eqref{zet0} and \eqref{fexpan} as the
effect of small instantons).

On the other hand, when $\kappa \to\infty$ (and $P,Q$ are fixed)
the potential term in \eqref{diffa} becomes large at all real $x$,
and therefore the Wronskian in \eqref{zet0} can be found by
straightforward application of WKB technique. This results in the
asymptotic $\kappa\to\infty$ series  \bea\label{irexpa}
Z_{\theta = 0}(\, {\bf P}\,
|\, \kappa\, ) \simeq {\rm g}_{D}^2\
\exp\bigg\{-\sum_{l=1}^{\infty}\, {{I_{2l-1}({\bf
P})}\over{\kappa^{2l-1}}}\, \bigg\}\,, \eea where
$I_{2l-1} ({\bf P}) = I_{2l-1}(P,Q)$ are
certain polynomials in $P^2,$ and $Q^2$ of the degree $l$, with
coefficients which depend only on $n$. It turns out that these
polynomials are exactly the vacuum eigenvalues of the local IM of
the ``paperclip series'' $\{{\mathbb I}_{2l-1}\}$, \bea\label{shsy}
{\mathbb I}_{2l-1}\ |\, {\bf P}\,\rangle = R^{1-2l}\
 {I_{2l-1}(P,Q)}\
|\,{\bf P}\,\rangle
\eea 
(see \cite{LVZ} for details). This
form of IR expansion suggests that at $\theta=0$ the IR fixed
point is just the Dirichlet boundary condition for the free field
${\bf X} = (X,Y)$: \bea\label{dirichlet} (X_B, Y_B) = (0,0)\,.
\eea According to\ \eqref{irexpa} the RG flow approaches the
Dirichlet fixed point along irrelevant direction which is a
combination of densities of the paperclip IM $\{{\mathbb
I}_{2l-1}\}$.

In this paper we extend the proposal of \cite{LVZ} to the case
$\theta=\pi$. The starting point is the same differential
equation\ \eqref{diffa}, but now instead of $\Psi_{+}$ in
\eqref{zet0}, we take another solution which grows as
$\re^{-\kappa x}$ at large negative $x$. Of course, this condition
alone does not define the solution uniquely since, besides overall
normalization, one can always add any amount of $\Psi_{+}(x)$.
Usually it is difficult to make unambiguous definition of a
growing solution, but in our case the following property of\
\eqref{diffa} helps. Let us consider $x$ as a complex variable.
The potential $V(x)$ is analytic function of $x$ with the
branching-point singularities at all points where $\re^x$ turns to
$-1$. Let us make brunch cuts from each of the points $x=\pi\ri\, (2
N+1)$, $N=0,\pm 1, \pm 2, \ldots$ to $+\infty$ parallel to the
real axis, and choose the branch of $V(x)$ on which $(1+\re^x)^n$
is real and positive on the real axis of the $x$-plane. Now,
restricting attention to the domain $\Re e\, x < 0$ one finds that
the potential $V(x)$ has the periodicity property
\bea\label{vperiod} V(x+2\pi\ri) = V(x) \qquad (\Re e\, x < 0)\,.
\eea Consequently, the equation\ \eqref{diffa} has two Bloch-wave
solutions ($2\kappa\not\in {\mathbb  Z}$): \bea\label{blochwaves}
\Psi_{\pm}(x+2\pi \ri) = \re^{\pm 2\pi{\rm
i}\,\kappa}\,\Psi_{\pm}(x) \ \ \qquad (\Re e\, x < 0)\,, \eea where
the Bloch factors are found by taking the limit $\Re e\, x \to
-\infty$. At this point we assume that $2\kappa$ is not an
integer, so that the conditions\ \eqref{blochwaves} specify the
two independent solutions $\Psi_{\pm}(x)$ uniquely, up to their
normalizations. Of course, the solution $\Psi_{+}(x)$ defined this
way decays as $\re^{\kappa x}$ at $\Re e\, x \to -\infty$, and the
asymptotic condition\  \eqref{psiplusdef} also fixes its
normalization. The solution $\Psi_{-}(x)$ grows at large negative
$\Re e\,  x$, and its normalization can be fixed by specifying the
leading asymptotic in this domain. Thus we define $\Psi_{-}(x)$ by
the conditions \bea\label{psiminusdef} &&\Psi_{-}(x + 2\pi\ri) =
\re^{-2\pi{\rm i}\,\kappa}\,\Psi_{-}(x) \quad (\Re e\, x < 0)\,,\\
&& \Psi_{-}(x) \to {{\re^{-\kappa x}}\over{\Gamma(1-2\kappa)}}\  \
\ \  \quad {\rm as} \quad \ \ \ \  \Re e\, x \to -\infty\,.
\nonumber \eea {} It is possible to show that both $\Psi_{+}(x)$
and $\Psi_{-}(x)$ defined by \ \eqref{psiplusdef} and \
\eqref{psiminusdef} are entire functions of $\kappa$, and
\bea\label{psiana}\Psi_{-}(\kappa\, |\, x) = \Psi_{+}(-\kappa\, |\, x)\,,
\eea
where we temporarily exhibited the dependence of $\Psi_{\pm}$ of
the parameter $\kappa$. From the definitions \eqref{psiplusdef}
and \eqref{psiminusdef} we have \bea\label{wpm}
W\big[\Psi_{-},\Psi_{+}\big] = {{\sin (2\pi\kappa)}\over{\pi}}\,.
\eea Our proposal for $Z_{\theta=\pi}$ is \bea\label{zetpi}
Z_{\theta=\pi}(\, {\bf P}\, |\, \kappa) = {\rm g}_{D}^2\,
\sqrt{\pi}\ \Big({{2\kappa}\over
\re}\Big)^{-2\kappa}\,W\big[\Xi, \Psi_{-}\big]\,. \eea

The $\kappa\to 0$ form of the Wronskian in\ \eqref{zetpi} can be
derived through the perturbative evaluation of the solutions
$\Xi(x)$ and $\Psi_{-}(x)$, as it was done in\ \cite{LVZ} for the
Wronskian in \ \eqref{zet0}. This leads to the expansion similar
to \ \eqref{uvzeto}, \bea\label{uvzetpi} Z_{\theta=\pi}(\,P,Q\,|\,
\kappa\,) = B(P,Q)\ F_{\theta=\pi}(P,Q\, |\, \kappa) + B(-P,Q)\
F_{\theta=\pi}(-P,Q\, |\, \kappa)\,, \eea where $B(P,Q)$ is the
same as in \ \eqref{uvzeto}, while \bea\label{fpiexpan}
F_{\theta=\pi}(P,Q|\kappa) \simeq
\kappa^{-2\kappa}\,\sum_{i,j=0}^{\infty}\,f_{i,j}(P,Q)\,(-\kappa)^i\,
\kappa^{{{2j}\over n}}\,, \eea with exactly the same coefficients
$f_{i,j}(P,Q)$ as in\ \eqref{fexpan}. This form follows from the
fact that integer powers of $\kappa$ in\ \eqref{fpiexpan} come
from perturbative expansion of the solution $\Psi_{-}(x)$; in view
of\ \eqref{psiana} they are related to corresponding integer
powers in\ \eqref{fexpan} by the change of the sign, $\kappa\to
-\kappa$. At the same time, the powers of $\kappa^{2\over n}$
appear as the result of expansion of $\Xi(x)$, and hence they
remain unchanged in\ \eqref{fpiexpan} as compared to
\eqref{fexpan}. Since the powers of $\kappa^{2\over n}$ are
interpreted as the perturbative contributions to the functional
integral \eqref{fint}, while the integer powers of $\kappa$ are
due to the instanton contributions, the form 
\eqref{uvzetpi},\,\eqref{fpiexpan}
is exactly what one expects to have from the
definition \eqref{topsum} at $\theta=\pi$. This property of the UV
expansion was the main motivation of our proposal \eqref{zetpi}.

Accepting \eqref{zetpi}, we can address the problem of the IR
behavior of the paperclip model at $\theta=\pi$. One has to find
the $\kappa\to\infty$ asymptotic of the Wronskian in
\eqref{zetpi}. Unlike the case $\theta=0$, Eq.\,\eqref{zet0},
this turns out to be rather subtle problem. While at $\kappa\to
\infty$ the WKB approximation still formally applies to
\eqref{diffa}, the problem is analogous to the problem of finding
the ``over-the-barrier'' reflection amplitude in quantum mechanics
which in the semiclassical approximation requires identifying
appropriate ``turning points'' (the zeroes of the potential
$V(x)$) in the complex $x$-plane \cite{Landau}. In our case, at
large $\kappa$ all the complex turning points approach the
singular points $\re^x=-1$ where the semiclassical approximation
for\ \eqref{diffa} breaks down. Finding correct $\kappa\to\infty$
expansion of \eqref{zetpi} requires analysis of the solutions of
\eqref{diffa} in the vicinities of the singular points $\re^x=-1$.
The calculations are rather involved, and will be present them
elsewhere. The result is the asymptotic $\kappa\to\infty$
expansion 
\bea\label{irexpi} Z_{\theta = \pi}(\, {\bf P}\, |\,
\kappa\,) \simeq {\rm g}_{D}^2\ \, \re^{a\,\kappa}\ \, T(\,{\bf P}\,
|\,\kappa)\ \exp\bigg\{\, \sum_{l=1}^{\infty}\, {{I_{2l-1}({\bf
P})}\over{\kappa^{2l-1}}}\,  \bigg\}\,, \eea where $a$ is a
constant \footnote{ $a=2\gamma_{E}+2\psi\big(1+{\textstyle{n\over
2}}\big)$ with $\psi(z)={\textstyle { {\rm d}\over {\rm d} z}}\,
\log\Gamma(z),\ \gamma_{E}=-\psi(1)$.}, $I_{2l-1}({\bf P})$ are the
same eigenvalues of the ``paperclip IM'' ${\mathbb I}_{2l-1}$ as in
\eqref{irexpa}, and $T(\, {\bf P}\, |\, \kappa\, )$ is the
asymptotic series in inverse powers of $\kappa^{2\over{n+2}}$,
\bea\label{tseries} T(\, {\bf P}\, |\, \kappa\,) \simeq
2\cos\big({\textstyle {\pi\,Q\over \sqrt{n+2}}}\big) +
\sum_{l=1}^{\infty}t_{l}(P,Q)\ \kappa^{-{{2l}\over{n+2}}}\,. \eea
Here again $P$ and $Q$ are the $X$ and $Y$ components of the
zero-mode momentum, i.e. ${\bf P} = \big(P,Q)$, and the
coefficients $t_l(P,Q)$ are in principle computable through
perturbative solution of \eqref{diffa} in the vicinity of the
singular points $\re^x=-1$. The first two coefficients can be 
evaluated in closed form. We have \bea\label{tone} &&t_1 (P,Q) =
\bigg({{n+2}\over 2}\bigg)^{2\over{n+2}}\,
{{\Gamma\big({1\over 2}-{1\over {n+2}}\big)}\over{\sqrt{\pi}\,\Gamma
\big(1-{1\over{n+2}}\big)}}\,\times\\
\nonumber&&\bigg[{{n+2}\over{n+4}}-{{(n+2)\,Q^2-n\,P^2-
1}\over{2\,((n+2)\,Q^2-1)}}\bigg]\,{{2\,\pi^2}\over{\Gamma
\big({Q\over\sqrt{n+2}}-{1\over{n+2}}\big)\, \Gamma\big(-{Q\over\sqrt{n+2}}-
{1\over{n+2}}\big)}} \,,
\eea
while the expression for $t_2 (P,Q)$ is somewhat cumbersome, and we
present it in  Appendix.  

The most important difference of\  \eqref{irexpi} from
\eqref{irexpa} is in the factor \eqref{tseries} which involves the
powers of ${\kappa^{-{2\over{n+2}}}}$; its presence indicates that
in this case the IR fixed point differs from the trivial Dirichlet
boundary constraint\ \eqref{dirichlet}. The series \eqref{tseries}
starts with the term $2\cos\big({\textstyle{ \pi\,Q\over
\sqrt{n+2}}}\big)$ which suggests that in the limit $\kappa \to
\infty$ the boundary values of ${\bf X}$ are constrained to two
points, \bea\label{irconstraint} (X_B, Y_B) = \big(0, \pm
{\textstyle {\pi\over\sqrt{n+2}}}\big)\,.
\eea
Eq.\,\eqref{irconstraint} characterizes the IR fixed point of the paperclip
boundary flow at $\theta=\pi$. At large but finite $\kappa$ the
boundary value $\big(X_B, Y_B\big)$ is allowed to jump between the two points
\eqref{irconstraint}, the possibility of such jumps being responsible
for the higher-order terms in the series\ \eqref{tseries}. In other words,
the RG flow arrives at the IR fixed point \eqref{irconstraint} along
certain irrelevant boundary fields which generate such jumps. Obvious
candidates are the boundary vertex operators
\bea\label{paravertex} V_{\pm}(\tau) = [\partial_{\sigma}X]_B \
\re^{\pm {{{\rm i}\,{\tilde Y}_B }\over\sqrt{n+2}}}(\tau)\,, \eea
where $\tau$ is natural coordinate along the boundary (we write
$z/R = \re^{{\sigma + {\rm i}\tau}\over R}$, so
that the boundary $|z|=R$ is at $\sigma=0$), $[\partial_{\sigma}X]_B =
\partial_{\sigma}X(\sigma,\tau)|_{\sigma=0}$ is the normal derivative
of the field $X$ at the boundary, and ${\tilde Y}_B (\tau)$ is the
boundary value of ${\tilde Y}$, the T-dual of $Y$
\footnote{The T-dual of the free massless field is defined as usual,
through the relations
  $\partial_{\tau}{\tilde Y} = \ri\, \partial_{\sigma} Y,$ and $
\partial_{\sigma}{\tilde Y} = -\ri\, \partial_{\tau} Y$.}.
The exponentials $\exp\big({\pm {{\ri\,{\tilde
Y}_B }\over\sqrt{n+2}}}\big)(\tau)$ create jumps in the boundary value
$Y_B$ at the point $\tau$, i.e. the boundary values $Y_B$ to the right
and to the left of each vertex $V_{\pm}(\tau)$ differ by the amount
$\pm {2\pi\over\sqrt{n+2}}$. Note that the dimensions of the fields
\eqref{paravertex} equal to $1 + {1\over{n+2}}$, which is in exact
agreement with the fractional power $\kappa^{-{2\over{n+2}}}$ appearing
in the expansion \eqref{tseries}. According to these arguments, it seems
likely that the full asymptotic series \eqref{tseries} can be
generated by expanding the following
expression \bea\label{pexpo} &&\ \ \ \ \ \ \ T(\, {\bf P}\, |\, \kappa\, ) =\\
&&   \big\langle\, {\bf P} \, \big|\, {\rm Tr}\Big[\re^{ {{ {\rm i}\pi
Q}\over{ \sqrt{n+2}}}\,\sigma_{z}}\ {\cal T}\exp\Big\{ \lambda
\int_{0}^{2\pi R}\rd\tau\,\big(\sigma_{-}V_{+}(\tau) + \sigma_{+}
V_{-}(\tau)\big)\Big\} \Big]\, \big|\, B_D \,\big\rangle\,. \nonumber \eea
Here $|\, B_D \,\rangle$ is the boundary state of the free theory
\eqref{baction} with the Dirichlet boundary condition\
\eqref{dirichlet} which is the superposition $|\, B_D \,\rangle =
\int\,\rd^2{\bf P}\ |\, I_{\bf P}\,\rangle$ of the Dirichlet
Ishibashi states $|\, I_{\bf P}\,\rangle$. In what follows we assume that
the exponentials in \eqref{paravertex} are normalized in such a way that
\bea\label{vertexnorm} \langle\,{\bf P}\, |\,  V_{+}(\tau)V_{-}(\tau')\,
|\,  B_D\,\rangle \to 2\ |\tau-\tau'|^{-2-{2\over{n+2}}} \quad
{\rm as} \quad \tau-\tau' \to 0\,. \eea
The $\tau$-ordered
exponential is understood in terms of its expansion in powers of
the parameter $\lambda$ (which will be related to the parameter
$E_{*}$, see Eq.\,\eqref{lambdae} below),
\bea\label{expoo}&&T(\, {\bf P}\, |\, \kappa\, ) =
2\cos\big({\textstyle {\pi\,Q\over \sqrt{n+2}}}\big) +
{\textstyle {{\lambda^2}\over 2}}\ \int_{0}^{2\pi R}\rd\tau
\int_{0}^{\tau}\rd\tau'\times
\\ &&\ \ \ \ \big\langle\, {\bf P}\, \big|
\, \Big[\, \re^{-{{{\rm i}\pi Q}\over{\sqrt{n+2}}}}\
V_{+}(\tau)\, V_{-}(\tau') +
\re^{{{{\rm i}\pi Q}\over{\sqrt{n+2}}}}\
V_{-}(\tau)\, V_{+}(\tau')\,
\Big]\, \big|\, B_D \,\big\rangle + \ldots\,.\nonumber
\eea
The relations $\sigma_{\pm}^2 =0$ guarantee that the boundary values
of $Y$ are limited to two points indicated in \eqref{irconstraint}. With
this understood, two eigenstates of $\sigma_z$ with the eigenvalues
$\pm 1$ appear to be in correspondence with two possible boundary
values $Y_B = \pm {\pi\over\sqrt{n+2}}$. At the same time, the
$\sigma$-matrices in\ \eqref{pexpo} can be regarded as operators
representing an additional boundary degree of freedom; then the expression
in the exponential in \eqref{pexpo} is interpreted as the boundary action
describing perturbation of the Dirichlet boundary condition. Anyhow,
the $g$-factor at the IR fixed
point (i.e. \eqref{irexpi} evaluated at $\kappa=\infty$ and ${\bf P}=0$)
is $2\, {\rm g}_{D}^2$, not ${\rm g}_{D}^2$ as it was in the case of
the $\theta=0$ IR fixed point \eqref{dirichlet}.

As was mentioned above, the dimensions of the vertex operators
\eqref{paravertex} are greater than one. As the result, the
integrals appearing in the expansion \eqref{expoo} generally
diverge when $\tau$-separations between the insertions
$V_{\pm}(\tau)$ become small. One has to specify some
regularization to make the expression \eqref{pexpo} meaningful. We
assume here the ``analytic regularization'' common in conformal
perturbation theory. The prescription is to consider $n$ as an
arbitrary complex number and to evaluate the integrals in the
domain of $n$  where they converge (specifically, at $\Re e\, n <
-4$), and then to continue in $n$ to the real positive values. It
turns out that this continuation yields unique finite results for
all integrals involved in the expansion \eqref{expoo}. Thus, the
$\lambda^2$ term explicitly written in \eqref{expoo} is readily
evaluated in analytic form. Remarkably, its dependence of the
momenta $(P,Q)$ appears exactly the same as in \eqref{tone}, and
upon identification \bea\label{lambdae} \lambda =
{{\sqrt{n}}\over{\sqrt{2}\,\Gamma\big(-
{1\over{n+2}}\big)}}\,\bigg({{n+2}\over
{E_{*}}}\bigg)^{1\over{n+2}}\ , \eea 
it coincides with the first
($l=1$) term $t_1 (P,Q)\,\kappa^{-{2\over{n+2}}}$ of the expansion
\eqref{tseries}. We believe that under the identification
\eqref{lambdae}, and with the analytic regularization described
above, the expression \eqref{pexpo} reproduces all terms of the
asymptotic series \eqref{tseries}, i.e. it plays the role of the
``effective IR theory'' describing the approach of the
$\theta=\pi$ paperclip flow to its IR fixed point
\eqref{irconstraint}. Note that the effective theory \eqref{pexpo}
is expected to capture only the ``perturbative'' part
\eqref{tseries} of the full IR expansion \eqref{irexpi}, and it
does not seem to have any control over the ``nonperturbative''
terms due to the local IM in \eqref{irexpi}.

The relation \eqref{lambdae} is singular at $n=\infty$. Correspondingly,
obtaining correct $n\to\infty$ limit from the effective theory \eqref{pexpo}
is not exactly straightforward. At $n\gg 1$ alternative form of the
IR effective theory is more convenient. It is written as
\bea\label{pexpon} &&\ \ \ \ \ \ T(\, {\bf P}\, |\, \kappa\, ) =\\
&&  \langle\, {\bf P} \, |\, {\rm Tr}\, {\cal T}\exp\Big\{
\int_{0}^{2\pi R}\rd\tau\,\big(
{\textstyle {{\alpha_{\rm x}}\over 2}}\
\sigma_{x}\, [\partial_{\sigma} X]_B (\tau) +
{\textstyle{{\alpha_{\rm y}}\over
  2}}\
\sigma_{y}\,
[\partial_{\sigma} Y]_B(\tau)\big)\Big\} \, |\, B_D \,\rangle\,,
\nonumber 
\eea
where again $[\partial_{\sigma}X]_B$ and  $[\partial_{\sigma}Y]_B$ are the
normal derivatives of the fields $X$ and $Y$ at the boundary, and
$\alpha_{\rm x}$, $\alpha_{\rm y}$ are coupling constants.
Formally, \eqref{pexpon}
can be brought to the form \eqref{pexpo} by field-dependent gauge
transformation \bea\label{gauge} \sigma_{\rm a} \to
\re^{-{{{\rm i}\alpha_{\rm y}}\over 2}\, \sigma_{\rm y}{\tilde Y}_{B}}(\tau)\
\sigma_{\rm a}
\ \re^{{{{\rm i}\alpha_{\rm y}}\over 2}\, \sigma_{\rm y}
{\tilde Y}_{B}}(\tau)\ ,  
\eea
with obvious relabeling of the $\sigma$-matrices to match the
notations in \eqref{pexpo}. The subtlety in this transformation
is in the renormalization of the parameters. Perturbation theory
in the couplings $\alpha_{\rm x}$, $\alpha_{\rm y}$ in \eqref{pexpon} has
logarithmic UV divergences which lead to renormalization of these
parameters. In view of the $X \leftrightarrow Y$ symmetry the
corresponding RG flow equations can be written as
\bea\label{alphabeta}
-E\ {\rd g_{\rm x}
\over{\rd E}} =
\beta(g_{\rm x},g_{\rm y})\,, \ \ \ -
E\ {\rd g_{\rm y}\over{\rd E}} =
\beta(g_{\rm y},g_{\rm x})\,, \eea where $E$ is the RG energy scale, and
$g_{\rm x} = \alpha_{\rm x}^2$,\
 $g_{\rm y} = \alpha_{\rm y}^2$. The model
\eqref{pexpon} was studied in \cite{Affleck}, where the leading
(one-loop) term of the beta-function is presented,
\bea\label{hahahs} \beta(g_{\rm x}, g_{\rm y}) = -2\ g_{\rm x}
g_{\rm y} + O(g^3)\ . \eea The higher loop terms depend on the
renormalization scheme. One can note that \eqref{pexpo} is easily
evaluated in the limit when $g_{\rm x}\to 0$ with $g_{\rm y}$ kept
finite (or when $g_{\rm y} \to 0$ with finite $g_{\rm x}$). It
follows from this solution that a class of ``natural'' schemes
exists in which \bea\label{schemecond} \beta(g,0) = \beta(0,g)
=0\,, \ \ \ {\rd\over {\rd g}}\, \beta(g,h)\big|_{g=0} =-2\, h\,.
\eea In any such scheme \bea\label{twoloop} \beta(g_{\rm x},g_{\rm
y}) = -2\ g_{\rm x} g_{\rm y} + b\ g_{\rm x}^2 g_{\rm y} + O(g^4)\
, \eea where the constant $b$ is scheme independent. Explicit
two-loop calculation yields $b=4$. The terms $\sim g^4$ and higher
still depend on the scheme. It looks likely that any RG flow
equation of the form \eqref{alphabeta} with the $\beta$-function
satisfying \eqref{schemecond} and having the expansion
\eqref{twoloop} with $b=4$ can be brought, order by order in $g$,
to the following convenient form \bea\label{irrg} -E\ {\rd {\hat
g}_{\rm x}\over{\rd E}} = -E\ {\rd {\hat g}_{\rm y}\over{\rd E}} =
-{{2\,{\hat g}_{\rm x} {\hat g}_{\rm y}}\over{1+{\hat g}_{\rm
x}+{\hat g}_{\rm y}}}\ , \eea where ${\hat g}_{\rm x} = g_{\rm
x}/(1-g_{\rm x})$, ${\hat g}_{\rm y} = g_{\rm y}/(1-g_{\rm y})$,
by appropriate redefinition of the coupling constants $g_{\rm
x}$,\ $ g_{\rm y}$ (we have explicitly checked this statement up
to the order $g^6$). The RG flow \eqref{irrg} conserves the
difference $\epsilon = {\hat g}_{\rm y} -{\hat g}_{\rm x}$. In the
coordinates ${\hat g}_{\rm x}, {\hat g}_{\rm y}$ the RG
trajectories are straight lines which end at the IR fixed points
$g_{\rm x}=0$ if $\epsilon>0$, or $g_{\rm y}=0$ if $\epsilon < 0$.
Let us assume here $\epsilon \geq 0$. The RG invariant $\epsilon$
can be related to the parameter $n$ in \eqref{pexpo},
\bea\label{invariant} \epsilon \equiv {\hat g}_{\rm y} -{\hat
g}_{\rm x} = {1\over{n+1}}\ . \eea Indeed, at the IR fixed point
${g}_{\rm x}=0$ while $g_{\rm y} = {\textstyle {\epsilon\over
1+\epsilon}}$; the last quantity must coincide with the dimension
of the coupling constant $\lambda$ in \eqref{pexpo}. Integrating
\eqref{irrg} one finds \bea\label{rgsolve} {\hat g}_{\rm x} =
{{\epsilon\, \rho}\over{1-\rho}}\,, \ \ \ \ {\hat g}_{\rm y} =
 {{\epsilon}\over{1-\rho}}\,,
\eea
where
$\rho$ decays towards IR according to the equation \bea\label{rho}
{{\epsilon\,\rho^{{1\over 2\epsilon}}}\over{1-\rho}} = {{E\over
E_{0}}}\ ,
\eea
where $E_0$ is the integration constant; one can set $E_0 = C\,E_{*}$ with
an arbitrary constant $C$. The equations \eqref{rgsolve},\,\eqref{rho}
define a family of $E$-dependent renormalized coupling constants
${\hat g}_x, \ {\hat g}_y$. It is clear that \eqref{pexpon} admits
perturbative expansion in these coupling constants, and if one
chooses $E=R^{-1}$ the coefficients of this series are undependent
of $R$.

Let us consider the case $n=\infty$ which is of special interest
because it corresponds to the circular brane model\ \cite{SLAZ} at
$\theta=\pi$,  and hence is equivalent to the Ambegaokar-Eckern-Sch\"on 
model \cite{AES} of quantum dot at the degeneracy point.
Obviously, in this limit the form \eqref{tseries} of
the IR asymptotic expansion of the boundary amplitude is no longer
meaningful. Instead, in this case the factor $T(\, {\bf P}\, |\,
\kappa\, )$ in \eqref{irexpi} rather expands in powers the
``running coupling constant'' ${\hat g}$ related to $\kappa$ by
the equation \bea\label{isoscale} {\hat g}\ \re^{-{1\over{2{\hat
g}}}} = {1\over{\re^{c_0}\,\kappa}}\ . \eea The same relation is
obtained by integrating the RG equation \eqref{irrg} with ${\hat
g}_{\rm x} = {\hat g}_{\rm y} = {\hat g}$\,.
 It is possible to show that at $n=\infty$ the factor
$T(\, {\bf P}\, |\, \kappa\, )$ in \eqref{irexpi} expands as
\bea\label{isot} T(\, {\bf P}\, |\, \kappa) = 2 - (\pi{\bf P})^2\
{\hat g} + \big(\, \pi^2 - 2\,(\pi{\bf P})^2 +{\textstyle{{1\over
  12}}}\,(\pi{\bf P})^4\, \big)\ {\hat g}^2 + O({\hat g}^3)\ ,
\eea
where ${\hat g}={\hat g}(\kappa)$ is determined through 
\eqref{isoscale} with
\bea\label{shshy}
c_0 = 3+\gamma_E\ ,
\eea
which is in perfect agreement with the results of
renormalized perturbation theory in \eqref{pexpon}.

The differential equation \eqref{diffa} can be integrated
numerically, thus providing high precision numerical data for the
disk partition functions \eqref{zet0} and \eqref{zetpi}. To
handle the circular-brane case $n=\infty$ an appropriate limit of the
differential equation \eqref{diffa} has to be taken. Making shift
of the variable $x=y-\log n$ and then sending $n$ to infinity, one
obtains the differential equation (124) of  Ref.\,\cite{LVZ},
while the asymptotic condition for the solution $\Xi$ is also
suitably modified, see Eq.\,(126) of  Ref.\,\cite{LVZ}.

\begin{figure}[ht]
\centering
\includegraphics[width=10cm]{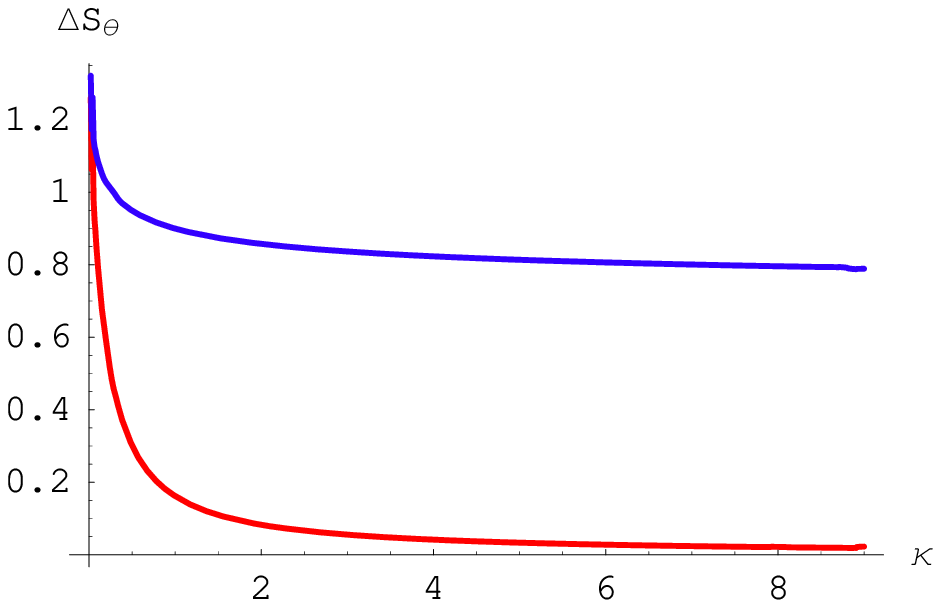}
\caption{The boundary entropy  $\Delta S_{\theta}=
S_{\theta}-\log ({\rm g}^2_D)$\
\eqref{entropy}\ for  the circular  brane model, as a function of
the dimensionless inverse temperature $\kappa$ for $\theta=0,\pi$,
derived from \eqref{zet0} and \eqref{zetpi}.} \label{fig-entrop}
\end{figure}

In Table \ref{table-ampl} we compare 
data for the boundary
entropy \cite{affleck,Friedan}  of the circular brane 
(Ambegaokar-Eckern-Sch\"on) model \bea\label{entropy}
S_{\theta}(\kappa) = \Big(\, 1-\kappa\,{\rd\over {\rd\kappa}} \,
\Big)\, \log Z_{\theta}(\, {\bf 0}\, |\, \kappa\, )\ , \eea
obtained by numerical integration of the differential equation
corresponding to $n=\infty$ with first few terms of the IR
expansions \bea\label{irterms} \Delta S _{\theta=0}^{{\rm IR}}(\kappa) 
&=& 
{1\over{6\, \kappa}} - {1\over{180\, \kappa^3}} +{71\over 105000\,
\kappa^5}+O(\kappa^{-7})\ ,
\nonumber \\
\Delta S _{\theta=\pi}^{{\rm IR}}(\kappa) 
&=&\log(2)+ {\textstyle
{\pi^2\over 2}}\, {\hat g}^2+ 2\pi^2\, {\hat g}^3-
{\textstyle
{\pi^2(24+\pi^2)\over 12}} \, {\hat g}^4+O({\hat g}^5)-\\ 
 &&{1\over{6\, \kappa}} + {1\over{180\, \kappa^3}} - {71\over
105000\, \kappa^5}+O(\kappa^{-7})\  ,\nonumber
\eea 
where $\Delta S _{\theta}\equiv S_{\theta}-\log ({\rm g}^2_D)$.
The  boundary entropy for
$\theta=0,\pi$ as a function of the dimensionless inverse
temperature $\kappa={E_{*}\over 2\pi T}$ is plotted in Fig.\,\ref{fig-entrop}.
\begin{table}[ht]
\begin{center}
\begin{tabular}{| r | l l | l l |}
\hline
\rule{0mm}{4mm}
$\kappa$\hspace{1mm}
& $\Delta S_{\theta=0}$ & $\Delta S_{\theta=0}^{{\rm IR}}$ & $\Delta S_{\theta=\pi}$
& $\Delta S_{\theta=\pi}^{\rm IR}$ \\
\hline
\hline
\rule{0mm}{3.6mm}
0.2  & 0.5877   & 2.25198    & 1.0264    & 0.02039 \\
0.3  & 0.4474   & 0.62806    & 0.9967    & 1.19216 \\
0.4  & 0.3607   & 0.39590    & 0.9660    & 1.18637  \\
0.5  & 0.3032   & 0.31053    & 0.9490    & 1.12614 \\
1.0  & 0.1617   & 0.16179    & 0.8996    & 0.97467 \\
1.5  & 0.1095   & 0.10955    & 0.8737    & 0.92069 \\
2.0  & 0.0826   & 0.08266    & 0.8574    & 0.89149 \\
2.5  & 0.0663   & 0.06632    & 0.8453    & 0.87246 \\
3.0  & 0.0554   & 0.05535    & 0.8366    & 0.85876 \\
3.5  & 0.0475   & 0.04749    & 0.8291    & 0.84827 \\
4.0  & 0.0416   & 0.04158    & 0.8231    & 0.83989 \\
4.5  & 0.0370   & 0.03698    & 0.8179    & 0.83298 \\
5.0  & 0.0333   & 0.03329    & 0.8136    & 0.82716 \\
5.5  & 0.0303   & 0.03027    & 0.8097    & 0.82216 \\
6.0  & 0.0278   & 0.02775    & 0.8064    & 0.81780 \\
6.5  & 0.0256   & 0.02562    & 0.8034    & 0.81396 \\
7.0  & 0.0238   & 0.02379    & 0.8006    & 0.81053 \\
7.5  & 0.0222   & 0.02221    & 0.7980    & 0.80745 \\
8.0  & 0.0208   & 0.02082    & 0.7959    & 0.80466 \\
8.5  & 0.0196   & 0.01960    & 0.7938    & 0.80212 \\
\hline
\hline
\end{tabular}
\end{center}
\caption{Comparison of the boundary entropy
$\Delta S_{\theta=0,\pi} = S_{\theta=0,\pi}-\log ({\rm g}^2_D)$ for the 
circular brane as computed from the differential equation
with
few terms of the IR asymptotic expansions 
$\Delta S^{\rm IR}_{\theta=0,\pi}$ given in 
Eqs.\,\eqref{irterms}.}
\label{table-ampl}
\end{table}

\begin{figure}[ht]
\centering
\includegraphics[width=10cm]{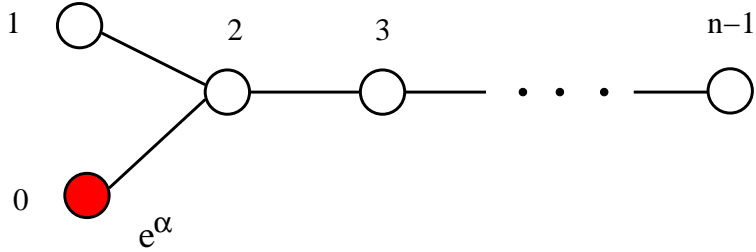}
\caption{Incidence diagram for the TBA system in the case
of the paperclip brane with an integer
$n\geq 3$
and $P=0,\ Q=\pm{1\over\sqrt{n+2}}$. The source term indicated near the
corresponding node.}
\label{fig-tba}
\end{figure}

Omitting details, we would like to mention here that using methods 
developed in \cite{Voros} (see
also \cite{blzz, dooorey}) one can reduce the calculation of the
Wronskians \eqref{zet0},\,\eqref{zetpi}\ at special values of the
parameters to solution of certain system of nonlinear integral
equations, the ``TBA system''. For example, for integer $n$ and
\bea\label{sanshgx} P=0\, ,\ \ \ \ \ Q=\pm{1\over \sqrt{n+2}}\, ,
\eea the amplitude $Z(\, {\bf P}\,|\, \kappa)$\ \eqref{zdef}\ is
expressed through solution of the  $D_n$-type TBA system,
associated with the diagram in Fig.\,\ref{fig-tba}. It is
remarkable that this system coincides exactly with the system of
TBA equations for the Toulose limit of the $n$-channel Kondo model
\cite{tsvel}, while the entropy $S^{\rm (Kondo)}_j$ of the 
impurity spin  $j={n-1\over 2}$ is given by
\bea\label{bssgtg}
S^{\rm (Kondo)}_{n-1\over 2}=
\Big(\, 1-\kappa\,{\rd\over {\rd\kappa}} \,
\Big)\, \log\big(\, {\rm g}_D^{-4}\ Z_{\theta=0}(\, {\bf P}^*\, |\, \kappa\,)\ 
Z_{\theta=\pi}(\, {\bf P}^*\, |\, \kappa\,)\, \big)\ ,
\eea
where ${\bf P}^*=\big( 0\, ,\, {\textstyle{1\over \sqrt{n+2}}}\, \big)$
and $ E_*$ is identified  with the Kondo temperature.
Notice that at this choice of $P,Q$ the
coefficient $t_1$ in
the low-temperature expansion of the
partition function  (\ref{tseries}) vanishes and
the expansion starts with $t_2$ (see Table \ref{table-numt} in Appendix).
Also, this  TBA system\ (see Fig.\,\ref{fig-tba})\ differs only in the structure of
the ``source terms'' from the TBA system associated with the
$H_{n}^{(0)},\, H_{n}^{(\pi)}$ models \cite{FZ}, 
which are  certain integrable
perturbations of the minimal ${\mathbb Z}_n$-parafermionic CFT. 

At $n=\infty$, the  boundary entropy of the
circular brane model with $\theta=0,\, \pi$ is given by
\bea\label{sssj} \Delta S_{\theta=k\pi}(\kappa)=
\int_{-\infty}^{\infty}{\rd\beta\over 2\pi}\ \re^{\gamma-\beta}\ \
{\log\big[1+\re^{-\varepsilon_{k }(\beta)} \big]\over
\cosh^2(\beta-\gamma)}\, \ \ \ \ \ \ \ (k=0,\, 1)\  , \eea where
$\gamma=\log(4\pi\kappa)$ and $\varepsilon_{k }, \ k=0,1,2,
\ldots$ solve the infinite chain of integral equations:
\bea\label{TBA}
0&=&\varepsilon_k(\alpha)+\int_{-\infty}^{\infty}{\rd\beta\over
2\pi}\ {\log\big[\big(1+\re^{-\varepsilon_{k+1}(\beta)}\big) \,
\big( 1+\re^{-\varepsilon_{k-1}(\beta)}\big)\big] \over
\cosh(\beta-\alpha)}+\nonumber\\ && \delta_{k,2}\
\int_{-\infty}^{\infty}{\rd\beta\over 2\pi}\ {\log\big[
1+\re^{-\varepsilon_{0}(\beta)}\big]\over \cosh(\beta-\alpha)} \ \
\ \  (k=2,\, 3,\, 4\ldots\infty)
\ ,\\
\re^{\alpha}\ \delta_{k,0}&=&\varepsilon_k(\alpha)+
\int_{-\infty}^{\infty}{\rd\beta\over 2\pi}\
{\log\big[1+\re^{-\varepsilon_{2}(\beta)}
\big]\over \cosh(\beta-\alpha)}\ \ \ \ (k=0,\, 1)\ .\nonumber
\eea

\bigskip

To sunmmarize, we have proposed exact partition function of 
a nontrivial model of boundary 
interaction -- the paperclip model. 
There are parallels between this model on one side, 
and the bulk sigma-models such as the $O(3)$ nonlinear sigma 
model and the ``sausage'' model \cite{sausage} on the other.
In particular, according to our proposal, the addition of the 
topological term with 
$\theta = \pi$ in the paperclip model 
leads to a non-trivial IR fixed point. In the limit 
when the parameter $n$ goes to infinity the paperclip
model reduces to the ``circular brane'', which is equivalent to
the Ambegaokar-Eckern-Sch\"on model of a quantum dot. At finite
integer $n$ our proposal suggests intriguing formal relation to the 
$n$-channel Kondo model in the Toulose limit.

\bigskip

\section*{Acknowledgments}

The authors grateful to Lev B. Ioffe, Mikhail V. Feigel'man
for valuable discussions
and   Dmitri M. Belov
for help with
graphical software.
SLL and ABZ are grateful to Vladimir V. Bazhanov,
Vladimir A. Fateev and  Alexei B. Zamolodchikov
for sharing their insights. AMT is 
grateful to Igor  L.  Aleiner for discussions on quantum dots. 

\bigskip

\noindent

The research of SLL  and ABZ is supported
in part by DOE grant $\#$DE-FG02-96 ER 40949.
SLL also acknowledges the support from the
Institute for Strongly Correlated and Complex System at BNL,
where most of the results of this paper have been
obtained.
AMT  acknowledges the support from
US DOE under contract number $\#$DE-AC02-98 CH 10886.
ABZ gratefully acknowledges hospitality and generous support from
Foundation de l'Ecole Normale Sup${\acute {\rm e}}$rieure.

\bigskip
\bigskip
\section*{Appendix}
Here we present an integral representation
for the second coefficient $t_2$ appeared in the expansion
\eqref{tseries}, which was obtained by an analytic technique  developed
in Appendix A of Ref.\cite{oscel}. Bellow we use the following notations
\bea\label{mdndh}
\xi={1\over n+2}\ ,\ \ \ p=\sqrt{n}\ \ {P\over 2}\, ,\ \ \ \ 
q=\sqrt{n+2}\ \ {Q\over 2}\ .
\eea
The coefficient $t_2$ can be written as
\bea\label{ssjsa}
t_2=-4\, \sin^2(\pi\xi)\cos\big(2\pi\xi(q+1)\big)\, 
 a^2_1-
4\, \sin(2\pi\xi)
\sin\big(2\pi(q+1)\xi\big)\, a_2\, .
\eea
Here
\bea\label{nsshhy}
a_1=(2\xi)^{-2\xi}\,
{ \Gamma(\xi)\Gamma({1\over 2}-\xi)\over
2\, \sqrt{\pi}
}
\, \bigg[\, {1-2\xi\over 2(1+2\xi)}+
{p^2\over 2\,  (q^2-{1\over 4})}\,
\bigg]\,
{\Gamma(1+(2q+1) \xi )\over \Gamma((2q-1) \xi)} ,
\eea
and  $a_2$ in \eqref{ssjsa} admits the following
representation for ${\Re e}\, q>0$
and  $0<\xi<{1\over 2}\ \big(\xi\not={1\over 4}\big)$:
\bea\label{aaasjsuau}
a_2={\Gamma\big(1+2\xi(q+1)\big)\over \Gamma\big(2\xi(q-1)\big)}\ \Big(\, 
{A\over q^2-1}+B+C\, \Big)\ ,
\eea
where
\bea\label{skj}
A&=&-(2\xi)^{-4\xi}\ {\Gamma(1+2\xi)\, \Gamma({1\over 2}-2\xi)\over
8\xi\  \sqrt{\pi}}\ \Big(p^2+{1\over 4}\Big)^2+\nonumber \\ &&
(4\xi)^{-4\xi}\ { \sqrt{\pi}\Gamma({1\over 2}
-2\xi)\Gamma^2(1+3\xi)
\over 4\xi\, \Gamma^2({1\over 2}+\xi)\Gamma(1+2\xi)}\
\bigg[{1-2\xi\over 2(1+2\xi)}+{2p^2\over 3}\, \bigg]^2\ ,\\
B&=&(2\xi)^{-4\xi} \ {\Gamma(1+2\xi)\, \Gamma({5\over 2}-2\xi)\over
8\xi\, \sqrt{\pi} \, (1+4\xi)}\ ,\nonumber\\
C&=&
{(2\xi)^{-4\xi}\over 8 
\Gamma^2(1-\xi)\Gamma^2({1\over 2}+\xi)}\times \nonumber\\
&&\int_{-\infty}^{\infty}
{\rd\tau\over 2\pi}\, {F(\tau)\, \sinh(2\pi\tau)
\over (\tau+\ri q\xi)(\tau-\ri\xi)
(\tau+\ri \xi)}\,
\bigg[\, {1-2\xi\over 2(1+2\xi)}-{2\xi^2\, p^2\over 4\tau^2+\xi^2}
\, \bigg]^2\, ,\nonumber
\nonumber
\eea
with
$$F(\tau)=\Gamma(1-2\xi+2\ri\tau)\Gamma(1-2\xi-2\ri\tau)
\Gamma^2(1+\xi+2\ri\tau)\Gamma^2(1+\xi-2\ri\tau)\ . $$
Notice that  for $\xi= {1\over 4}\ (n= 2)$ one has
to use the L'${\hat{\rm o}}$pital's rule to 
evaluate $a_2$ from Eq.\,\eqref{aaasjsuau}.
Using \eqref{ssjsa}  it is straightforward to study the
large $n$-behavior $(\xi\to 0)$ of $t_2$:
\bea\label{soio}
&&t_2=2 \pi^2\ \xi^2\ \big[\, \xi\, \re^{2+\gamma_E}\,
\big]^{-4\xi} \, \Big[\, 1-2 q^2-2p^2-(3-4 q^2+4p^2)\ \xi+\\
&&\Big\{\, 4+{\textstyle{\pi^2\over 24}} -
 \pi^2\, \big(\, q^2-2\, q^4-{\textstyle{2\over 3}}\, p^4-
\big({\textstyle{8\over \pi^2}}-{\textstyle {1\over 3}}
\big)\,  p^2-{8\over 3}\, q^2\, p^2 \, \big)\Big\}\,
\xi^2+ O(\xi^3)\, \Big]\, .\nonumber
\eea

The above representation is very useful in numerical evaluation of 
this coefficient because of the fast convergency of the integral in
\eqref{skj}.
In Table 2 we present the numerical  values of the coefficient $t_2$ at
some integer $n$ and $P,\ Q$ given by\ Eq.\,\eqref{sanshgx}. 
These numbers were also confirmed  by numerical integration
of the TBA system depicted in   Fig.\,\ref{fig-tba}.
\begin{table}[ht]
\begin{center}
\begin{tabular}{| r | l| }
\hline
$ n$
&  $\quad t_2$    \\
\hline
\hline
\rule{0mm}{3.6mm}
1   & 0  \\
2   & ${\sqrt{2}\over 48}$   \\
3   & $0.0546105$  \\
4   & $0.0661040$\\
5   & $0.0683646$\\
6   & $0.0658731$\\
7   & $0.0613178$\\
8   & $0.0561029$\\
9   & $0.0509101$\\
10  & $0.0460445$\\
\hline
\hline
\end{tabular}
\end{center}
\caption{ Numerical values of the second coefficient $t_2$ in the expansion
\eqref{tseries} for $P=0,\ Q=\pm {1\over \sqrt{n+2}}$. }
\label{table-numt}
\end{table}

\end{document}